\pdfoutput=1
\documentclass[10pt,conference]{IEEEtran}
\normalsize
\newcommand{\subparagraph}{}
\usepackage{algorithmic}
\usepackage{algorithm}
\usepackage{amsmath}
\usepackage{amssymb}
\usepackage[tight,footnotesize]{subfigure}
\usepackage{subfigure}
\usepackage{graphicx}
\usepackage{multirow}
\usepackage{mathtools}
\usepackage{lipsum}
\usepackage{titlesec}
\usepackage{epstopdf}
\usepackage{mwe}
\usepackage{float}
\usepackage[utf8]{inputenc} 
\usepackage[english]{babel}
\usepackage [autostyle, english = american]{csquotes}
\MakeOuterQuote{"}

\usepackage[makeroom]{cancel}
\titlespacing\section{6pt}{10pt plus 4pt minus 4pt}{4pt plus 6pt minus 6pt}
\titlespacing\subsection{8pt}{10pt plus 4pt minus 4pt}{2pt plus 4pt minus 4pt}
\titlespacing\subsubsection{0pt}{12pt plus 2pt minus 2pt}{0pt plus 2pt minus 2pt}
\newcommand\numberthis{\addtocounter{equation}{1}\tag{\theequation}}

\allowdisplaybreaks

\begin{document}

\title{On Asymptotic Analysis of Zero-Delay Energy-Distortion Tradeoff Under Additive White Gaussian Noise}

\author{Ceren Sevinç, Ertem Tuncel \\ 
Department of Electrical and Computer Enginering, University of California, Riverside, CA. \\
Email: csevinc@ee.ucr.edu, ertem.tuncel@ucr.edu}

\IEEEtitleabstractindextext{%
\begin{abstract}
Asymptotic energy-distortion performance of zero-delay communication scenarios under additive white Gaussian noise is investigated. Using high-resolution analysis for quantizer design, the higher-order term in the logarithm of the distortion (termed the {\em energy-distortion dispersion}) is optimized while keeping the leading term (i.e., {\em energy-distortion exponent}) at its optimal value. For uniform and Gaussian sources, significant gains are observed compared to na\"{i}vely performed quantization, i.e., aimed at optimizing the source coding performance instead of the end-to-end distortion in joint source-channel coding.
\end{abstract}

\begin{IEEEkeywords}
Companding, energy-distortion dispersion, energy-distortion tradeoff, high-resolution quantization theory, joint source-channel coding, zero-delay.
\end{IEEEkeywords} }

\maketitle

\IEEEdisplaynontitleabstractindextext

\IEEEpeerreviewmaketitle

\section{Introduction}

Consider the communication scenario where a very slowly varying source is transmitted over an energy-limited additive white Gaussian coise (AWGN) channel. Due to the nature of the source, the channel can be utilized with a relatively high bandwidth compared to the source. At the same time, however, block coding of the source incurs an intolerable delay even for short block lengths. Therefore, the ideal transmission scheme should encode each source sample into a very long channel word. One application for this type of scenario is smart-grid systems in which a smart-meter measurement is taken every 15 minutes to be transmitted immediately to the central unit \cite{smart}. 

To make it amenable to analysis, we idealize this scenario such that the source is independent and identically distributed (i.i.d.), and each source sample is mapped separately into and infinite-length channel word. In this setup, it was argued in \cite{tuncel1} that the \textit{energy-distortion exponent}, defined as
\[
\Theta = \lim_{\gamma  \rightarrow \infty} -\frac{1}{\gamma} \ln D(\gamma) \; ,
\] 
i.e., the rate of decay of the minimum mean square-error (MSE) $D$ as the energy-to-noise ratio (ENR) $\gamma$  approaches infinity, provides a suitable performance measure, especially in the absence of a fully characterized energy-distortion tradeoff $D(\gamma)$. 

In \cite{tuncel1}, it was shown that the same energy-distortion exponent (i.e., $\Theta=1$) for an infinite-delay transmission of a Gaussian source over an AWGN channel can be achieved under the current scenario of zero-delay transmission, provided outage events with arbitrarily small probability are allowed. 

In this paper, we take a different approach and analyze the energy-distortion exponent for the {\em overall} MSE, without recourse to conditioning on the non-outage event. We also pursue a more detailed characterization in the form of
\begin{equation}
\label{pursuit}
-\ln D(\gamma) = \Theta\gamma + \Upsilon(\gamma) + o(1)
\end{equation}
for large $\gamma$, where $\Upsilon(\gamma)$ is sub-linear in $\gamma$, i.e., 
\[
\lim_{\gamma\rightarrow\infty}\frac{\Upsilon(\gamma)}{\gamma}=0 \; .
\]
Seeing a parallel between \eqref{pursuit} and recent results in finite blocklength source and channel coding, whereby higher-order terms of the coding rate as a function of the blocklength $n$ is investigated~\cite{kostina,yuri}, we define the higher order term $\Upsilon(\gamma)$ as the {\em energy-distortion dispersion}.

In pursuit of finding $\Theta$, the maximum possible exponent, Burnashev \cite{burnashev} arrived at the conclusion that for uniform sources, 
\[
- \ln D(\gamma) \leq \frac{1}{6}\gamma + C \ln (1+\gamma) -\ln C
\]
for some constant $C$ and large enough $\gamma$. This implies $\Theta\leq\frac{1}{6}$ as an upper bound to the maximum energy-distortion exponent. This bound is in fact tight as implied by  \cite{knopp} and \cite{merhav}, which showed $\Theta\geq\frac{1}{6}$ through achievable schemes. Therefore, for uniform sources, Burnashev's result implies the following upper bound on the dispersion:
\[
\Upsilon(\gamma) \leq C \ln (1+\gamma) -\ln C \; .
\]

\begin{figure*}
\centering
\includegraphics[width=0.7\textwidth]{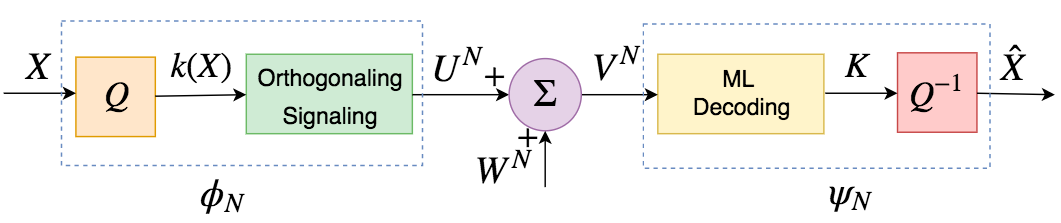}
\caption{Block diagram of the proposed coding scheme.}
\label{fig:coding}
\end{figure*}

The main contribution of this work can be summarized as follows. We show that $\Upsilon(\gamma)$ is lower bounded by a constant. Towards that end, we devise a joint-source channel coding scheme based on quantization followed by orthogonal signaling and maximum-likelihood (ML) decoding, and employ high-resolution quantization theory to analyze the resultant MSE.
We then take a step further and tighten this lower bound on $\Upsilon(\gamma)$ by maximizing the constant.
This entails finding the {\em point-density function} for the high-resolution quantizer that would minimize end-to-end MSE distortion taking into account the channel decoding errors, as opposed to na\"{i}vely using the quantizer that would minimize the source coding MSE.
As a result of this optimization, we obtain $0.383$dB and  $0.0943$dB improvement in the distortion for Gaussian and uniform sources, respectively, compared to quantization optimized solely for source coding performance. It is worth mentioning that the methodology that we suggest can be easily applied to other bounded and unbounded source distributions as well. For the Gaussian source, we also compare our result with \cite{knopp}, where uniform quantization with a bounded domain was used. While their approach also yields the same exponent, namely $1/6$, their dispersion diverges as the ENR increases without bound, thereby bringing about considerable degradation in MSE for high ENR values.

The rest of the paper is organized as follows. The system model is introduced in Section \ref{model}.  We provide the details of the asymptotically optimal quantizer design for a Gaussian source in Section \ref{design}, and for a uniform source in Section \ref{design2}. Finally, in Section \ref{results}, we illustrate the simulation results and compare our results with previous work.

\section{System Model} \label{model}

Let $X$ be a real-valued scalar source to be transmitted over the channel
\begin{equation}
V^N = U^N + W^N
\end{equation}
where $U^ N$ and $V^ N$ are the channel input and output, respectively. The channel noise $W^N$ is independent of $U^N$ and  $W^N \sim \mathcal{N} ( \mathbf{0, \sigma}^2_W \mathbf{I}_N)$, where $\mathbf{\sigma}^2_W$ is  the noise variance and $\mathbf{I}_N$ is $N$-dimensional identity matrix. The encoder
\begin{equation*}
 \phi_{N}:\mathbb{R}\rightarrow \mathbb{R} ^{N}
\end{equation*}
\noindent maps $X$ into $U^N$, and the decoder
\begin{equation*} 
\psi_{N}:\mathbb{R}^{N} \rightarrow \mathbb{R} 
\end{equation*}
estimates $X$ as $\hat{X}$. 
The energy expended at the channel input is constrained as 
\[
|| U^N ||^2 \leq E
\]
and the reconstruction quality is measured by the usual square-error distortion
\[
D = \mathbb{E}[(X-\hat{X})^2] \; .
\]
We refer to $\frac{E}{\sigma^2_W}$ as the {\em energy-to-noise ratio} (ENR) and use the notation 
\[
\gamma=\frac{E}{\sigma^2_W} \; .
\]

We will focus on schemes where the source $X$ is first quantized using $N$ levels and the quantization index $k(X)$ is mapped into orthogonal channel input vectors $U^N$ such that
\begin{equation*}
U_t =  \begin{cases}
    \sqrt{E} ,& t=k(X) \\
    0,     &  t\neq  k(X)
\end{cases}
\end{equation*}
thereby enforcing $|| U^N ||^2 = E$.  At the receiver, $k(X)$ is decoded using maximum likelihood (ML) estimation as $\hat{K}$, and the source is reconstructed as the $\hat{K}$th quantization level. Occasional decoding errors will be denoted by the outage event
\begin{equation}
\mathcal{O} = \left\{ k(X) \neq \hat{K} \right\}.
\end{equation}
This proposed coding scheme is illustrated as a block diagram in Fig. \ref{fig:coding}. 

One convenient feature of ML estimation is that given the event ${\cal O}$, the reconstruction $\hat{X}$ is distributed {\em uniformly} over the {\em incorrect} reconstruction values of the quantizer.
Also, it is not difficult to see that $X$ is independent of ${\cal O}$. 

Using the outage notation, one can write the MSE as
\begin{eqnarray}
\lefteqn{\mathbb{E}[(X - \hat{X} ) ^2]} \nonumber \\ & = & \Pr [\mathcal{O}] \mathbb{E}[(X  \! - \!  \hat{X} )^2| \mathcal{O}]  \! +  \!  \Pr [\mathcal{O}^c] \mathbb{E}[(X \!  -  \! \hat{X} )^2| \mathcal{O}^c] \nonumber \\
\label{dist1}
& \leq & \Pr [\mathcal{O}] \mathbb{E}[(X  \! - \!  \hat{X} )^2| \mathcal{O}]  \! +  \!  \mathbb{E}[(X \!  -  \! \hat{X} )^2| \mathcal{O}^c]
\end{eqnarray}
In~\cite{tuncel1}, it was shown that while keeping the outage event at a vanishingly small probability, i.e., $\Pr[{\cal O}]\leq \epsilon$ for arbitrary $\epsilon>0$, one can ensure that 
\begin{equation}
\label{BestExponent}
-\frac{1}{\gamma} \ln \mathbb{E}[(X \!  -  \! \hat{X} )^2| \mathcal{O}^c]  \longrightarrow 1
\end{equation}
as $\gamma\rightarrow\infty$.
The significance of (\ref{BestExponent}) is that it coincides with the best exponent theoretically achievable in the Shannon-theoretic scenario of encoding infinitely long source blocks at once.  

In this work, we tackle (\ref{dist1}) in its entirety and investigate the behavior of $-\ln \mathbb{E}[(X - \hat{X} ) ^2] $ as a function of $\gamma$. No matter how small $\Pr[{\cal O}]$ is, it may affect the total expected distortion dramatically. More specifically, if it decays slower than $\mathbb{E}[(X \!  -  \! \hat{X} )^2| \mathcal{O}^c]$ as $\gamma\rightarrow\infty$, it will dominate (\ref{dist1}) and adversely affect the energy-distortion exponent.

\begin{figure}[h!]
\centering
\includegraphics[width=0.45\textwidth]{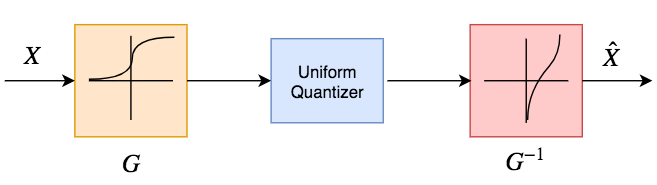}
\caption{Compander model, where $G$ is a nonlinear compressor and $G^{-1}$ is a nonlinear expander.}
\label{fig:comp}
\end{figure}

In our analysis, we employ tools from high-resolution quantization theory. The high-resolution assumption is justified by the fact that the number of quantization levels $N$ must increase exponentially with $\gamma$ to ensure an exponentially decaying distortion, as will be apparent in the sequel.
Distortion in the high-resolution regime is best understood with the help of companders~\cite{gersho} as shown in Fig. \ref{fig:comp}. First, a nonlinear compressor $G$ reduces the spread of large amplitudes and maps the source sample to $[0,1]$.  Then, the source is uniformly quantized in the compressed domain with $N$ levels. Finally, a nonlinear expander $G^{-1}$ reverses this process by expanding the small amplitudes of uniformly quantized output. The end-to-end effect becomes a non-uniform quantizer. In fact, any non-uniform quantizer can be put into this form~\cite{gersho}.

The point density function, $\lambda(x)=\frac{dG}{dx}$, provides one with an equivalent framework. It also has the convenient property that $\lambda(x)\geq 0$ and
\[
\int_{-\infty}^\infty \lambda(x)dx = 1 \; .
\]
That is, its behavior is the same as that of a probability density function (pdf).

It is well-known (cf.\ \cite{zador,gersho2}) that 
\begin{equation}
\label{BennettIntegral}
\lim_{N \rightarrow \infty} N^2 D(N) = \frac{1}{12} \int \frac{f(x)}{\lambda^2(x)}  dx
\end{equation}
where $D(N)$ denotes the MSE distortion incurred by an $N$-level quantizer, and $f(x)$ is the source pdf.
The integral in (\ref{BennettIntegral}) is known as the Bennett integral.
We will use (\ref{BennettIntegral}) to write
\begin{equation}
\mathbb{E}[(X - \hat{X} )^2| \mathcal{O}^c]  \leq (1+ \epsilon') \frac{1}{12N^2} \int_{-\infty}^{\infty} \frac{f_X(x)}{\lambda^2(x)} dx \label{bennett}
\end{equation}
for any $\epsilon' > 0$ and large enough $N$.

We also have an upper bound on the outage probability
\begin{equation}
\Pr [\mathcal{O}] \leq
 \begin{cases}
    2e^{\left( \ln N- \frac{\gamma}{4} \right) }    ,& \ln N < \frac{\gamma}{8} \\
    2e^{-\frac{1}{2} \left( \sqrt{\gamma} - \sqrt{2 \ln N} \right)^2},     & \frac{\gamma}{8} \leq  \ln N \leq   \frac{\gamma}{2} ,
\end{cases} \label{Pe}
\end{equation}
where we refer the reader to \cite{tuncel1} and to the references therein for a detailed derivation of (\ref{Pe}).
We will use (\ref{BennettIntegral}) and (\ref{Pe}) in (\ref{dist1}) to properly select $N$ such that the end-to-end distortion decays with the maximum possible speed.

\section{Transmission of a Gaussian Source} \label{design}

Let $X\sim{\cal N}(0,1)$ and $N$ have the form of
\begin{equation}
N = ce^{\tau \gamma }
\end{equation}
where $c$ is a constant to be optimized and $\tau$ is to be picked later. Clearly, defining $N$ exponentially increasing in $\gamma$ is necessary to ensure that (\ref{BennettIntegral}) decays exponentially in $\gamma$.
Also observe that with this choice, even mediocre values of $\gamma$ will quickly drive $N$ to a very large number, thus justifying the high-resolution assumption.

Using (\ref{Pe}), this choice yields 
\begin{equation}
\label{PeExponent}
\lim_{\gamma\rightarrow\infty}-\frac{1}{\gamma}\ln\Pr [\mathcal{O}] \geq
 \begin{cases}
   \frac{1}{4}-\tau    ,& \tau < \frac{1}{8} \\
   \frac{1}{2} (1-\sqrt{2\tau})^2 & \frac{1}{8} \leq  \tau \leq   \frac{1}{2} .
\end{cases}
\end{equation}
Similarly, using (\ref{bennett}), we obtain
\begin{equation}
\label{DExponent}
\lim_{\gamma\rightarrow\infty}-\frac{1}{\gamma}\ln\mathbb{E}[(X - \hat{X} )^2| \mathcal{O}^c] \geq 2\tau \; .
\end{equation}
Therefore, according to (\ref{dist1}), the remaining task is to understand how the term $\mathbb{E}[(X - \hat{X} )^2| \mathcal{O}]$ behaves.

Let $\mathcal{R}_i$ and $\hat{x}_i$ denote the $i$th quantization region and the corresponding quantized value, respectively. Also define $\tilde{X}$ as the discrete random variable uniformly distributed over the $\hat{x}_i$ values, independent of $X$. We then estimate the resultant distortion by
\begin{align*}
\! \mathbb{E}[( \! X \!\! - \! \! \hat{X} )^2| \mathcal{O}] \! &= \! \sum_{i=1}^N \mathbb{E} \! \left[  (X \! -\! \hat{X})^2 \! \mid \! \mathcal{O}, X \! \in \! \mathcal{R}_i \right] \! \Pr \left[ X \! \in \! \mathcal{R}_i \! \mid \! \mathcal{O}\right]\\
& \stackrel{(a)}{=} \! \sum_{i=1}^N \mathbb{E} \! \left[  (X - \hat{X})^2 \! \mid \! \mathcal{O}, X \! \in \! \mathcal{R}_i \right] \! \Pr \left[ X \! \in \! \mathcal{R}_i  \right]\\
&= \sum_{i=1}^N \sum_{\substack{{j=1} \\
{j \neq i}}}^N \mathbb{E} [ (X- \hat{x}_j)^2 \mid \mathcal{O}, X \in \mathcal{R}_i, \hat{X} = \hat{x}_j ]  \\
& \qquad \quad \times \! \Pr [  \hat{X} = \hat{x}_j \mid \mathcal{O}, X \in \mathcal{R}_i ] \Pr [X \in \mathcal{R}_i ] \\
&\stackrel{(b)}{=} \!  \frac{1}{ N \! \! - \!1}  \! \sum_{i=1}^N \sum_{\substack{{j=1} \\
{j \neq i}}}^N \! \mathbb{E} [ (X\! \!- \! \hat{x}_j)^2 \! \! \mid \! \mathcal{O}, \! X \! \in \! \mathcal{R}_i, \! \hat{X} \! =  \! \hat{x}_j ]   \\
& \qquad  \quad  \times \Pr [X \in \mathcal{R}_i ] \\
&\stackrel{(c)}{=} \! \! \frac{1}{ N \! \! - \!1}  \! \sum_{i=1}^N \sum_{\substack{{j=1} \\
{j \neq i}}}^N \! \mathbb{E} [ (X\! \!- \! \hat{x}_j)^2 \! \! \mid \!  \! X \! \in \! \mathcal{R}_i]  \Pr [X \! \in \! \mathcal{R}_i ] \\
&\leq \! \! \frac{N}{N \!\! - \!\! 1} \frac{1}{N} \! \sum_{i=1}^N \sum_{j=1}^N \! \mathbb{E} [ (X\! \!- \! \!\hat{x}_j \!)^2 \! \! \mid \!  \! X \! \! \in \! \mathcal{R}_i] \! \Pr [X \! \! \in \! \mathcal{R}_i ] \\
&\stackrel{(d)}{=} \! \frac{N}{N-1} \mathbb{E} [ (X-\tilde{X} ) ^2]\\
&= \! \frac{N}{N-1} \left( \mathbb{E} [X^2] + \mathbb{E} [\tilde{X}^2] \right)\\
& = \! \frac{N}{N-1} \left( 1 + \mathbb{E} [\tilde{X}^2] \right) \numberthis \label{outage}
\end{align*}
\noindent where $(a)$ follows from the independence of the outage event $\mathcal{O}$ and $X$, $(b)$ from the fact that when outage happens, $\hat{X}$ is distributed uniformly on the \textit{incorrect} quantized values $\hat{x}_j $,  $j \neq i$, $(c)$ from the independence of $X$ from $\mathcal{O}$  and $\hat{X}$ when $X \in \mathcal{R}_i$ is given, and finally $(d)$ from the fictitious variable $\tilde{X}$ being uniformly distributed over all quantization levels $\hat{x}_j$.
Thus, as $N$ approaches infinity, we can write \eqref{outage} as
\begin{equation}
E[(X - \hat{X} )^2| \mathcal{O}] \leq 1+ \int_{-\infty}^{\infty} \tilde{x}^2 \lambda(\tilde{x}) d\tilde{x}. \label{outage2}
\end{equation}

We are now ready to pick $\tau$.
Since according to (\ref{outage2}), $E[(X - \hat{X} )^2| \mathcal{O}]$ is bounded by a constant, one can see that the overall distortion expression in (\ref{dist1}) becomes the sum of two exponential terms whose exponents are bounded as in (\ref{PeExponent}) and (\ref{DExponent}). 
Also, while the exponent of $\Pr[{\cal O}]$ is decreasing with $\tau$, that of $E[(X - \hat{X} )^2| \mathcal{O}^c]$ increases with the same. Thus, the problem becomes a $\max \min$ problem, and these two terms should have the same exponent in order to minimize the distortion. By carrying out the calculations, the optimum value of $\tau$ is found to be $\tau = \frac{1}{12}$,  leading to
\begin{align}
\Pr[{\cal O}] &  \leq P_e \stackrel{\Delta}{=} 2ce^{-\gamma/6} \\
N &= ce^{\gamma/12}.
\end{align}
\noindent As a result, the overall distortion expression in (\ref{dist1}) becomes 
\begin{align*}
D&\! \leq 2ce^{-\gamma/6} \left(  1+ \int _{-\infty}^{\infty} {x}^2 \lambda({x}) d{x}  \right) + \\
& \qquad \qquad\qquad  \qquad\qquad   \frac{1}{12N^2} \int_{-\infty}^{\infty} \frac{f_X(x)}{\lambda^2(x)} dx \\
&\!= e^{-\gamma/6} \! \left( 2c \! + \! \! \int_{-\infty}^{\infty} \!\!\! \left( 2c x^2 \lambda(x) \!+ \! \frac{1}{12 c^2} \frac{f(x)}{\lambda^2(x)} \right) dx  \right) \numberthis \label{Lag_eqn} \\
&\stackrel{\Delta}{=} f_0(c).
\end{align*}

While in typical quantizer design problems the optimal $\lambda(x)$, i.e., minimizing (\ref{BennettIntegral}), can be shown to be given by the well-known Panter-Dite formula, i.e., 
\[
\lambda(x) = \frac{f(x)^{1/3}}{\int f(x')^{1/3}dx'}
\]
which, for the Gaussian source we consider, coincides with ${\cal N}(0,3)$, the current form of end-to-end distortion in (\ref{Lag_eqn}) dictates a different solution.
Writing the new problem formally, we have
\begin{equation*}
\begin{aligned}
& \underset{\lambda(x)}{\text{minimize}}
& f_0 (c) \\
& \text{subject to}
& - \lambda(x) &\leq 0 \\
&&  \int_{-\infty}^{\infty} \lambda(x) dx &=1 \\
\end{aligned}
\end{equation*}
It is easy to check that this is a convex optimization problem. 
One can therefore write an equivalent Lagrangian 
\begin{equation*}
L = f_0(c) - \int_{-\infty}^{\infty} \alpha(x) \lambda (x) dx + \beta \int_{-\infty}^{\infty} \lambda(x)dx.
\end{equation*}
and solve the optimization problem by seeking a solution to the Karush-Khun-Tucker (KKT) conditions 
\begin{equation*}
\begin{aligned}
&& \frac{\partial L}{\partial \lambda(x)} &= 0 \\
&& - \lambda(x) &\leq 0 \\
& & \int_{-\infty}^{\infty} \lambda(x) dx &=1 \\
&& \alpha(x) &\geq 0 \\
&& \alpha(x) \lambda(x)&=0.
\end{aligned}
\end{equation*}

Taking the partial derivative above yields
 \begin{equation}
 \label{LL}
 \frac{\partial L}{\partial \lambda(x)}=P_e x^2-\frac{1}{6N^2}\frac{f(x)}{\lambda^3(x)} -\alpha(x)+\beta = 0 \; .
 \end{equation}
Rewriting (\ref{LL}), we have
\begin{equation}
\lambda(x)= \frac{f(x)^{1/3}}{ \left[6N^2 \left( P_e x^2- \alpha(x)+\beta \right) \right]^{1/3}}.  
\end{equation}
 
We attempt to find $ \lambda(x)$ that satisfies KKT conditions for the case where $ \alpha(x) =0$ and $ \beta \geq 0.$
Thus, $\lambda(x)$ becomes 
\begin{equation}
\lambda(x) =  \frac{1}{ 6^{1/3}c^{2/3}} \frac{f(x)^{1/3}}{\left( 2cx^2+\beta e^{\gamma/6} \right)^{1/3}}.  \label{lambda}
\end{equation}
To satisfy the KKT conditions, it remains to pick $\beta$ such that $\lambda(x)$ integrates to $1$.
Since \eqref{lambda} is decreasing in $\beta$, we can always find such a $\beta$ provided that $\int \lambda(x)dx \geq 1$ for $\beta=0$.
Following this logic and substituting $\beta=0$ in \eqref{lambda}, we observe that $c$ has to satisfy
\begin{align*}
 1 &\leq \frac{1}{ 12^{1/3}c} \int_{-\infty}^{\infty}  \frac{f(x)^{1/3}}{x^{2/3}} dx \\
 & = \frac{1}{ 12^{1/3} (2 \pi)^{1/6}c} \int_{-\infty}^{\infty} \frac{e^{-x^2/6}}{x^{2/3}} dx\\
&= \frac{1}{c} \left( \frac{2}{9 \pi } \right)^{1/6}  \times 3  \times6^{1/6} \times \Gamma \left(\frac{7}{6} \right).
\end{align*}
or
\begin{align*}
c & \leq  \left( \frac{2^2 3^5}{ \pi}  \right)   ^{1/6}  \Gamma \left(\frac{7}{6} \right)\\
&= 2.41269638 \stackrel{\Delta}{=} c_0 \numberthis \label{c}
\end{align*}

\begin{figure}[h]
\centering
\includegraphics[width=0.48\textwidth]{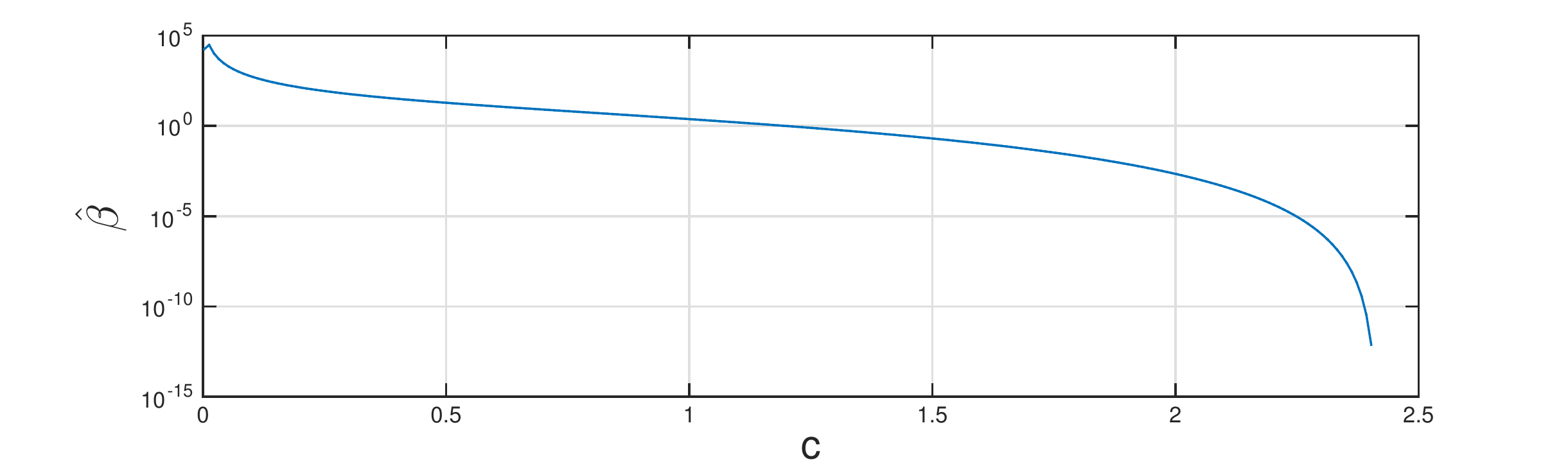}
\caption{$\hat{\beta}(c)$ vs $c$  for a Gaussian source.}
\label{fig:beta}
\end{figure}

For each $c  \leq c_0$, denote by $\beta(c)$ the value of $\beta$ satisfying  $\int \lambda(x)dx \geq 1$. For convenience, also set $\hat{\beta}(c)=\beta(c) e^{\gamma/6}$.
The relationship between $c$ and $\hat{\beta}(c)$ is shown in Fig. \ref{fig:beta}. 

Proceeding with \eqref{Lag_eqn},
\begin{align*}
D &  \leq e^{-\gamma/6} \left( 2c + \int_{-\infty}^{\infty} ( 2c x^2 \lambda(x)+ \frac{1}{12 c^2} \frac{f(x)}{\lambda^2(x)} dx ) \right) \\
&= e^{-\gamma/6} \left( 2c + \int_{-\infty}^{\infty}  \frac{ f(x) (\frac{4cx^2}{2cx^2 + \hat{\beta}(c)} +1)  }{12 c^2 \lambda^2(x)}  dx  \right) \\
&= e^{-\gamma/6} \left( 2c + \int_{-\infty}^{\infty}  \frac{ \lambda(x) (6cx^2 + \hat{\beta}(c))  }{2}  dx  \right) \\
& = e^{-\gamma/6} \left( 2c + 3c \int  x^2   \lambda(x)  dx + \frac{\hat{\beta}(c)}{2} \right) \\
&\stackrel{\Delta}{=} e^{-\gamma/6} \Omega (c) \numberthis \label{Omega}  \; .
\end{align*}

\begin{figure}[h!]
\centering
\includegraphics[width=0.47\textwidth]{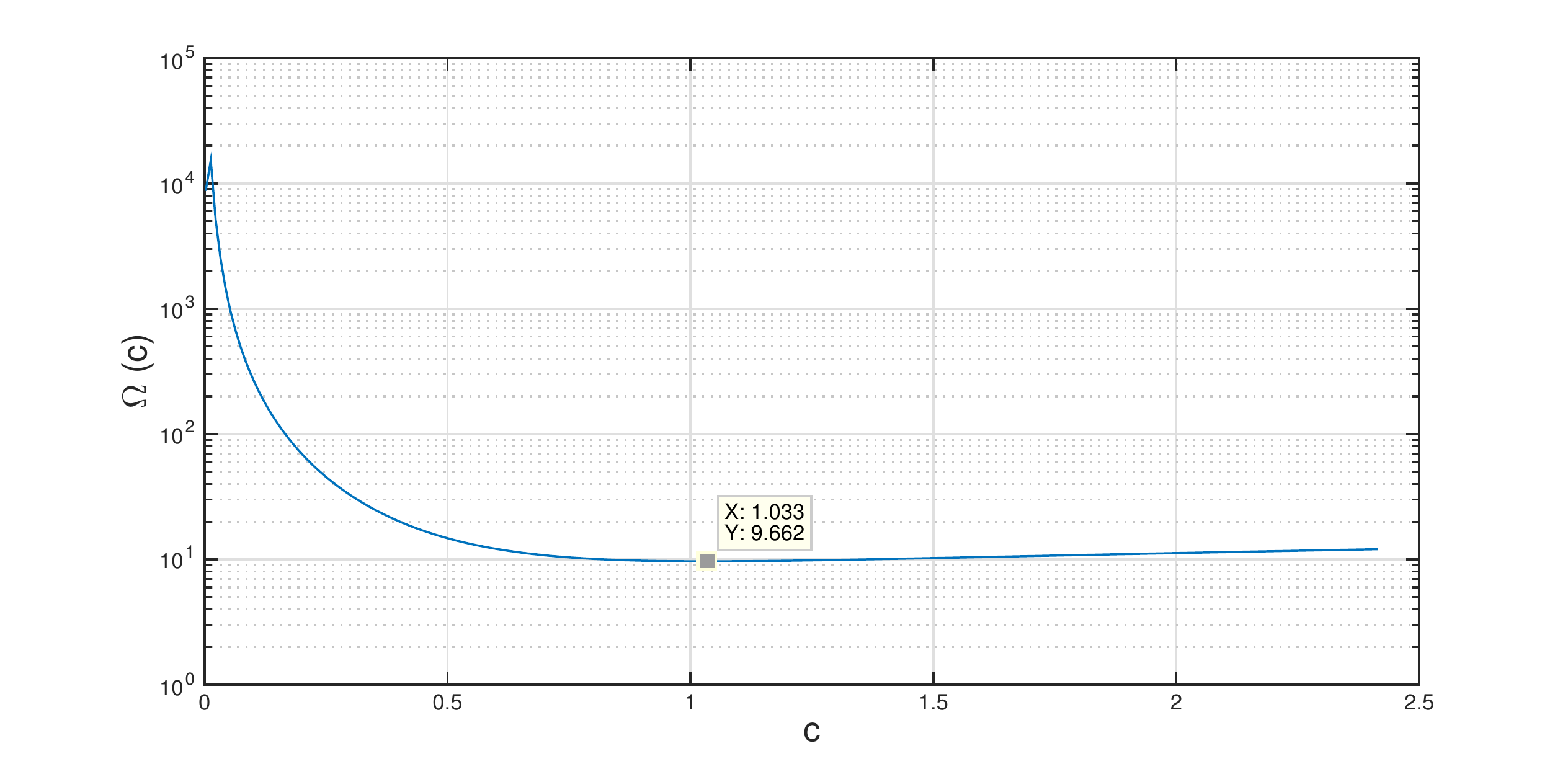}
\caption{$\Omega(c)$ vs $c$  for a Gaussian source.}
\label{fig:D}
\end{figure}

Fig. \ref{fig:D} depicts $\Omega(c)$ as a function of $c$. The optimal values of $c$, $\hat{\beta}(c)$, and $\Omega(c)$ are numerically found to be $c _{opt}\approx 1.0327$, $\hat{\beta}_{opt}\approx 2.0771$, and $\Omega_{opt}\approx 9.6622$, respectively. 

To analyze the characteristic of the distortion by means of higher order terms, let us rewrite \eqref{Omega} as
\begin{align}
- \ln D(\gamma) & \geq  \frac{1}{6}\gamma  - \ln \Omega_{opt}+ o(1)\nonumber \\
\label{dispersionformula}
 & \approx   \frac{1}{6}\gamma  - 2.2682 + o(1)
\end{align} 
for large $\gamma$.
Thus, for Gaussian sources, we obtained a lower bound to the dispersion, which is  $\Upsilon(\gamma) \geq -2.2682$.

\begin{figure}[h]
\centering
\includegraphics[width=0.47\textwidth]{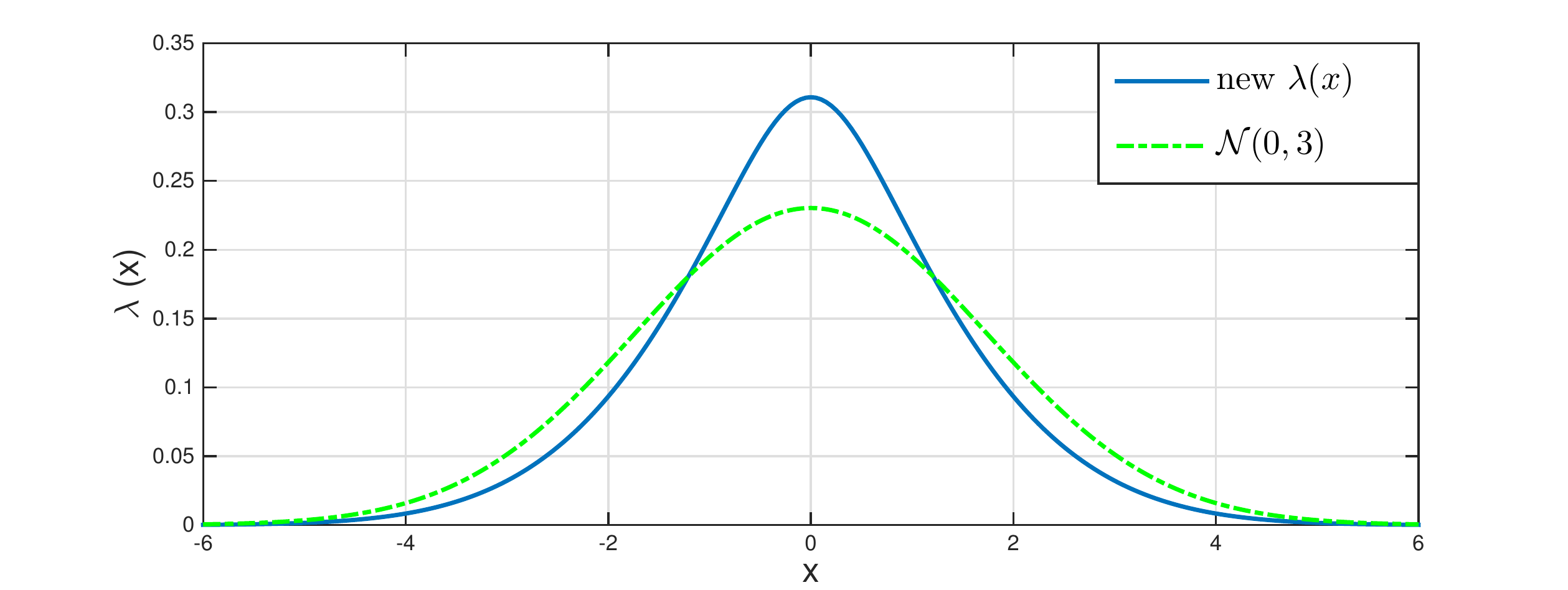}
\caption{Comparison between $\lambda(x)$ found and the na\"{i}ve point-density function $\mathcal{N}(0,3)$ for a Gaussian source.}
\label{fig:lambda}
\end{figure}

Fig. \ref{fig:lambda} shows the comparison between $\lambda(x)$ found and $\mathcal{N}(0,3)$, which would be the result of na\"{i}vely done quantization. It is not difficult to show that this choice (together with optimized $c$) would achieve a dispersion of $-2.3564$, and about $0.383$dB higher distortion for fixed $\gamma$.
 As can be observed from the figure, the new high-resolution quantizer has the point density function $\lambda(x)$ with a smaller variance (which is $1.93$) and it is no longer a Gaussian distribution.

\section{Transmission of a Uniform Source} \label{design2}
 
For the aforementioned scenario, we also provide performance results for a uniform source, which is frequently studied in the literature. When $X \sim \mathcal{U} [-1/2, 1/2]$, by following the same methodology, we derive the following upper bound:
\[
D \leq e^{-\gamma/6} \! \left( \frac{c}{6} \! + \! \! \int_{-\frac{1}{2}}^{\frac{1}{2}} \!\!\! \left( 2c x^2 \lambda(x) \!+ \! \frac{1}{12 c^2} \frac{1}{\lambda^2(x)} \right) dx  \right) \; .
\]
Minimizing this bound, we obtain the optimal $\lambda(x)$ as
\begin{equation}
\lambda(x) =  \frac{1}{6^{1/3}c^{2/3} \left( 2cx^2+\hat{\beta}(c)\right)^{1/3}}  \label{lambda2}
\end{equation}
for $ -1/2 \leq x \leq 1/2$.
By solving 
\[
1 = \int_{-1/2} ^{1/2} \lambda(x) dx 
\]
in the same fashion as before,  find $c_0=2.0801$, $\hat{\beta}_{opt}\approx 0.1385$, $c _{opt} \approx 0.8281$, and $\Omega_{opt} \approx 0.3884$. As a result, $-\ln D$ can be lower bounded as
\begin{eqnarray}
- \ln D(\gamma) & \geq  & \frac{1}{6}\gamma -\ln \Omega_{opt} +o(1) \nonumber \\
& \approx &  \frac{1}{6}\gamma + 0.9458 + o(1).
\end{eqnarray}
Hence, the energy-distortion dispersion for a uniform source is lower bounded as $\Upsilon(\gamma)\geq 0.9458$.

\begin{figure}[h]
\centering
\includegraphics[width=0.47\textwidth]{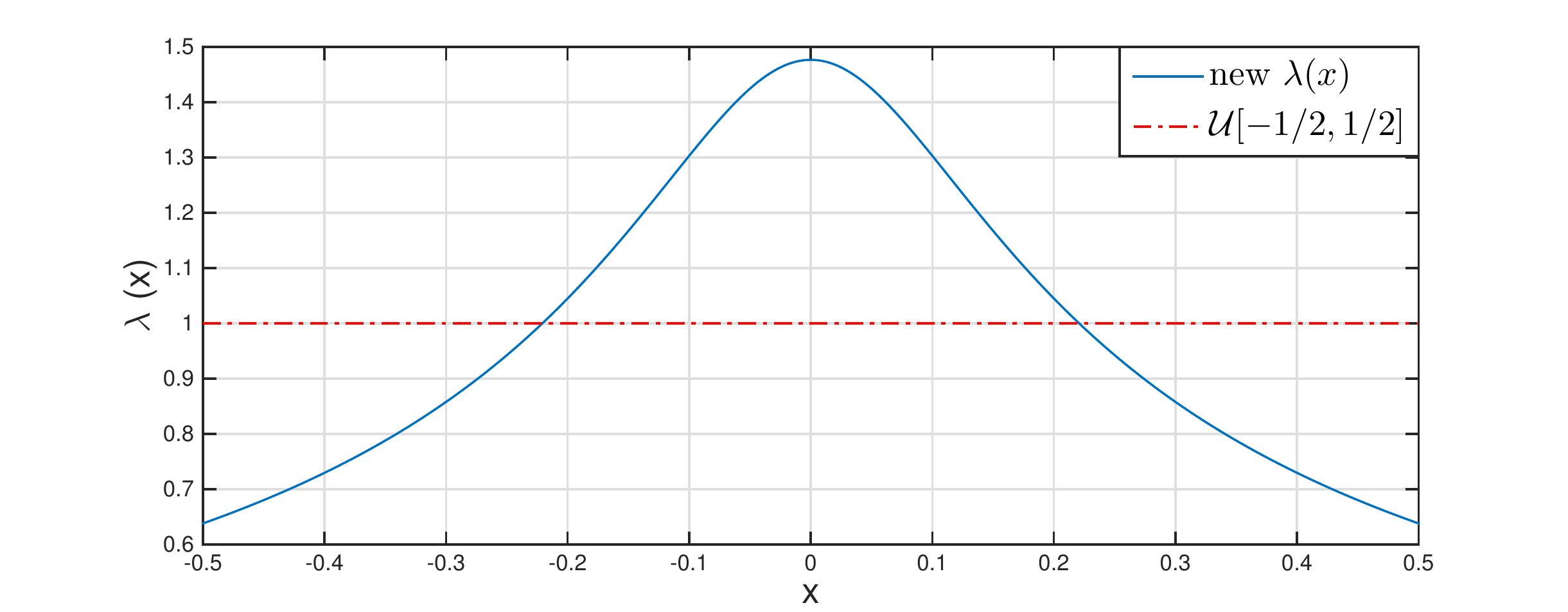}
\caption{Comparison between $\lambda(x)$ found and the na\"{i}ve point-density function $\mathcal{U}[-1/2,1/2]$  for a uniform source.}
\label{fig:lambda2}
\end{figure}

Fig. \ref{fig:lambda2} shows the comparison between the point-density functions  $\lambda (x)$ of our approach and the na\"{i}ve compander design, whcih would result in a uniformly distributed $\lambda (x)$.
It is not difficult to show that the uniform $\lambda(x)$ would achieve an energy-distortion dispersion of $0.9242$, and about $0.0943$dB higher distortion for fixed $\gamma$. 
 
\section{Comparison with Prior Work} \label{results}

In \cite{knopp}, an analytical upper bound for the distortion was derived using a linear quantizer and maximum a posteriori (MAP) receiver for a Gaussian source. Although their proposed scheme resulted in the same energy-distortion exponent under the asymptotic analysis, they disregarded the effect of the coefficient of the exponential term, causing a considerable performance degradation in the distortion, as we show next. By following the approach in \cite{knopp}, $D$ is calculated as follows.

\begin{align*}
D&= D_Q (1 - P_e) +D_e P_e \\
&<  D_Q  +D_e P_e\\
&< \frac{2e^{-\Delta ^2/2}}{\sqrt{2 \pi} \Delta} + \frac{\Delta^2}{(2^b-2)^2} + 2^{b\rho} e^{[ -\frac{\gamma}{2*} (\frac{\rho}{\rho+1})]} \times \\ 
& \qquad \qquad \qquad \left( 4 \Delta^2 + \frac{2(4 \Delta^2 +1)}{\sqrt{2 \pi } \Delta} e^{-\Delta^2/2}  \right)  \\
&= \frac{2^{-2b} }{\sqrt{2 \pi b \ln2}} +\frac{4b \ln2} {(2^b-2)^2}+  2^{b \rho-\frac{\gamma}{2\ln2} (\frac{\rho}{\rho+1})} \times \\
& \qquad \qquad \qquad \left( 16 b \ln 2 +\frac{(16 b \ln 2 +1)}{\sqrt{2 \pi b \ln2}} 2^{-2b} \right) 
\end{align*}
where $\rho$ is a constant between $[0,1]$, $\Delta =2 \sqrt{b \ln2} $ and $b \geq 2$ is an integer representing the quantization bits per source component.  For large enough $b$,  $\frac{1}{(2^b-2)^2}$ can be approximated by $2^{-2b}$, and $D$ becomes
\begin{align}
D \! &< \! 2^{-2b} \left( \frac{1}{\sqrt{2 \pi b \ln2}} +4b \ln2 \right) + 2^{\frac{\gamma}{2\ln2} (\frac{\rho}{\rho+1}) -b \rho} \times \nonumber \\
\label{KnoppDistortion}
& \qquad \qquad \qquad \left( 16 b \ln 2 +\frac{(16 b \ln 2 +1)}{\sqrt{2 \pi b \ln2}} 2^{-2b} \right) \; .
\end{align}

\noindent Hence, the largest energy-distortion exponent is calculated by
\begin{equation}
\theta= \max_{b,\rho} \min \left\{ 2b' \ln2, \frac{\rho}{\rho+1} \frac{1}{2} - \rho b' \ln2  \right\}
\end{equation}
where $b'= \frac{b}{\gamma}$. The optimal values are found to be $b_{opt} = \frac{\gamma}{12 \ln2}$ and $\rho_{opt}=1$ in the high ENR regime by leading to an optimal energy-distortion exponent, i.e., $\theta=\frac{1}{6}$. Thus, the upper bound on the distortion  becomes
\begin{align*}
 D< \! e^{-\gamma/6} \! \left( \! \frac{ \sqrt{ 6} }{ \sqrt{\pi \gamma}} \left(1\! + \! e^{-\gamma/6}  \! \left(\frac{4\gamma}{3} \! + \! 1\right)\! \right) \!  +\! \frac{5 \gamma}{3}  \! \right) \numberthis \label{Dknopp} \; .
\end{align*}
For large $\gamma$, this can be rewritten as
\[
-\ln D > \frac{1}{6}\gamma - \ln\frac{5\gamma}{3} + o(1)
\]
and thus the dispersion approaches  $- \infty$ as $\gamma\rightarrow\infty$. Our dispersion lower bound is clearly tighter. 

In \cite{knopp}, since the truncation was applied to a Gaussian source and it was quantized by using a uniform scalar quantizer, overload distortion was no longer negligible in terms of energy-distortion tradeoff. In \cite{hui}, it was shown that overall distortion for a Gaussian density under uniform quantization decreases as $\frac{\ln N}{ N^2}$ and overload distortion becomes asymptotically negligible. However, this causes the energy-distortion dispersion to diverge when there is a zero-delay constraint and $\gamma \rightarrow \infty$.

\begin{figure}[h]
\centering
\includegraphics[width=0.47\textwidth]{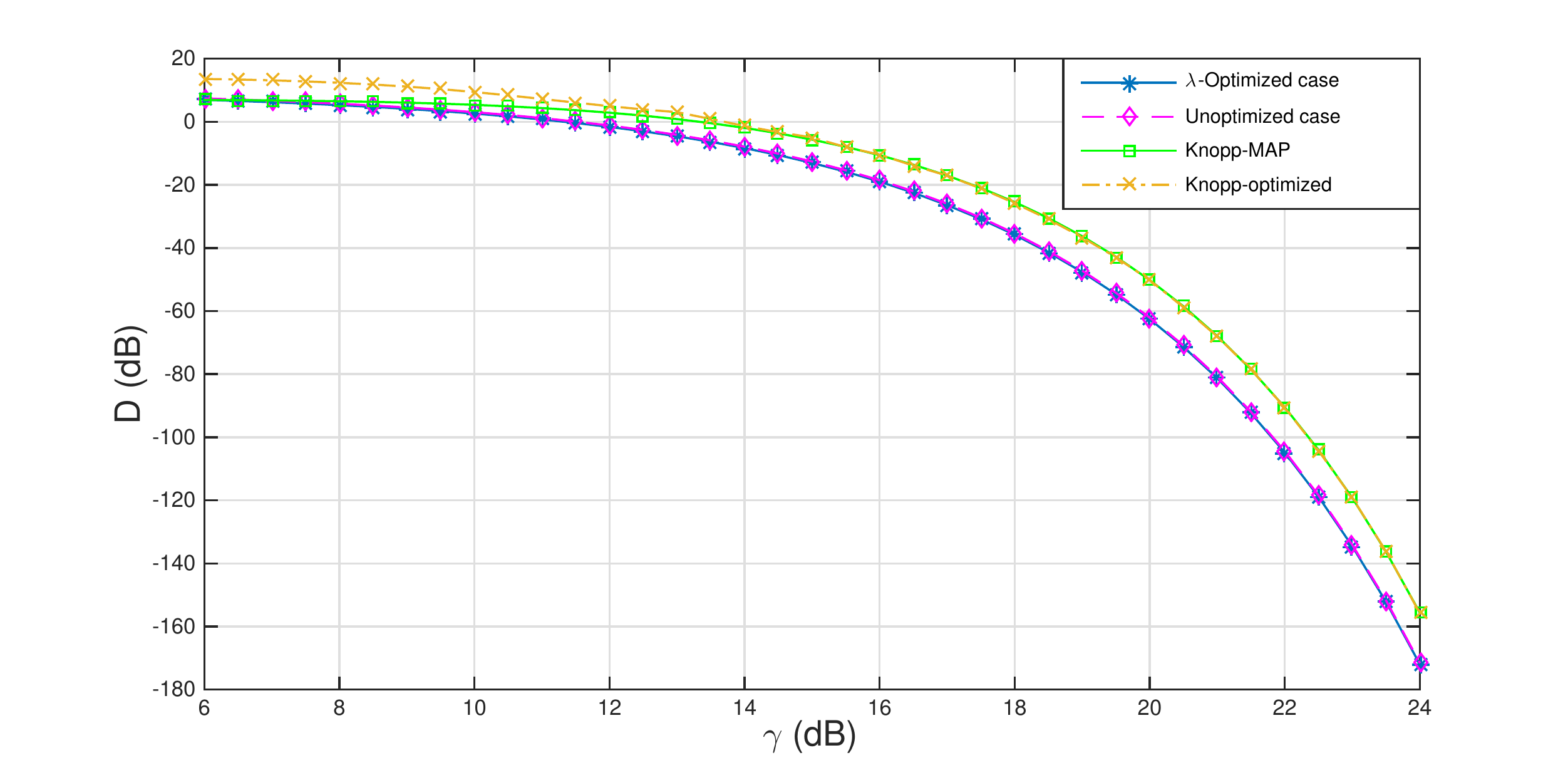}
\caption{Comparison of distortions in our work and \cite{knopp} for a Gaussian source.}
\label{fig:Dcomp}
\end{figure}

As an alternative to the analytical solution, we also provide a numerical solution to this problem by finding optimal values of $\rho$ and $b$ that minimize \eqref{KnoppDistortion} for each value of $\gamma$.  Comparison of the distortions of our approach, approaches in \cite{tuncel1}, \cite{knopp} is illustrated in Fig. \ref{fig:Dcomp}. The blue curve indicates the result of the distortion that is suggested in this work, in which $\lambda(x)$ is optimized, while the magenta curve shows the distortion for the case where $\lambda (x) = \mathcal{N}(0,3)$. The green curve depicts the numerical optimization of the distortion in \cite{knopp}, whereas the yellow curve shows the analytical result of the asymptotic distortion in that work. 
We note that the distortion gap between our approach and that of \cite{knopp}  diverges as $\gamma$ grows without bound.

\bibliographystyle{IEEEtran}
\bibliography{References}
 
\end{document}